%% file: paper.tex
\documentclass[sigconf]{acmart}

\usepackage[english]{babel}

\renewcommand\footnotetextcopyrightpermission[1]{} 
\setcopyright{none}

\usepackage{algorithm}
\usepackage{algorithmic}
\usepackage{multirow}
\usepackage{booktabs}
\usepackage{amsmath}
\usepackage{amsfonts}
\usepackage{graphicx}
\usepackage{textcomp}
\usepackage{subfigure}
\usepackage{graphics}
\usepackage{xspace}
 \usepackage{colortbl}

\newcommand{\modelname}{NetGPT\xspace}

\graphicspath{{./figs/}}
\DeclareMathOperator{\softmax}{softmax}
\acmDOI{}

\acmISBN{}

\acmPrice{}

\begin{document}
\title{NetGPT: Generative Pretrained Transformer for Network Traffic}

\author{Xuying Meng$^{1}$, Chungang Lin$^{1}$, Yequan Wang$^{2}$, Yujun Zhang$^{1}$}
\affiliation{
    \textsuperscript{\rm 1}Institute of Computing Technology, Chinese Academy of Sciences, China\\
    \textsuperscript{\rm 2}Beijing Academy of Artificial Intelligence, China\\
    \{mengxuying, linchungang22s, nrcyujun\}@ict.ac.cn, tshwangyequan@gmail.com}

\renewcommand{\shortauthors}{M. et al.}

\begin{abstract}
All data on the Internet are transferred by network traffic,  thus accurately modeling network traffic can help improve network services quality and protect data privacy. Pretrained models for network traffic can utilize large-scale raw data to learn the essential characteristics of network traffic, and generate distinguishable results for input traffic without considering specific downstream tasks. Effective pretrained models can significantly optimize the training efficiency and effectiveness of downstream tasks, such as application classification, attack detection and traffic generation. Despite the great success of pretraining in natural language processing, there is no work in the network field. Considering the diverse demands and characteristics of network traffic and network tasks, it is non-trivial to build a pretrained model for network traffic and we face various challenges, especially the heterogeneous headers and payloads in the multi-pattern network traffic and the different dependencies for contexts of diverse downstream network tasks.

To tackle these challenges, in this paper, we make the \textit{first} attempt to provide a generative pretrained model NetGPT for both traffic understanding and generation tasks. We propose the multi-pattern network traffic modeling to construct unified text inputs and support both traffic understanding and generation tasks. We further optimize the adaptation effect of the pretrained model to diversified tasks by shuffling header fields, segmenting packets in flows, and incorporating diverse task labels with prompts.
With diverse traffic datasets from encrypted software, DNS, private industrial protocols and cryptocurrency mining, expensive experiments demonstrate the effectiveness of our NetGPT in a range of traffic understanding and generation tasks on traffic datasets, and outperform state-of-the-art baselines by a wide margin. Code is available at \href{https://github.com/ict-net/NetGPT}{https://github.com/ict-net/NetGPT}.
\end{abstract}

\maketitle

\input{intro}

\input{related}

\input{core}

\input{experiment}

\section{Conclusions and Future Works} 
In this paper, we propose the first prertained model for both traffic understanding and generation tasks. We provide general encoding for multi-pattern network traffic with unified text inputs. We further optimize the adaptation effect of the pretrained model to diversified tasks by shuffling network fields, segmenting packets in flows, and incorporating diverse task labels with prompts. 
Expensive experiments demonstrate the effectiveness of our NetGPT on diverse traffic understanding and generation tasks and outperform state-of-the-art baselines by a wide margin.
Considering the great importance of network traffic, we believe our NetGPT will benefit a wide range of downstream network tasks and provide important theoretical and technical support for realizing networks intelligence.

One of the main limitations is the NetGPT is a relatively small model and the datasets are not large enough to cover all kinds of network traffic. 
For further work, we would like to enlarge the token size, parameter size and dataset size, and we believe the performance can increase with the size.
Additionally, we focus on software and algorithm improvement in this paper, and to decrease the computation delay, we would like to try if hardware like FPGA can increase the speed. Meanwhile, as we are the first work for network tasks, there are no benchmarks and the compared downstream tasks are limited, we would like to build the benchmark and explore more downstream tasks, such as generating complete traffic from scratch and observing how the generated traffic interacts with other hosts in real-world scenarios.

\bibliographystyle{ACM-Reference-Format}
\bibliography{reference}

\begin{table*}[h]
\centering
\caption{Traffic generation performance comparison.}
\label{tb:TG1}
\scalebox{0.7}{
\begin{tabular}{c|ccc|ccc|ccc|ccc|c}
\toprule
 \multirow{2}{*}{model} & \multicolumn{3}{c|}{ISXW}                                          & \multicolumn{3}{c|}{DoHBrw}                                        & \multicolumn{3}{c|}{USTCTFC}                                       & \multicolumn{3}{c|}{Cybermining}  & Avg                                 \\ \cline{2-14} 
                       & \multicolumn{1}{c|}{AC}     & \multicolumn{1}{c|}{F1}     & DR     & \multicolumn{1}{c|}{AC}     & \multicolumn{1}{c|}{F1}     & DR     & \multicolumn{1}{c|}{AC}     & \multicolumn{1}{c|}{F1}     & DR     & \multicolumn{1}{c|}{AC}     & \multicolumn{1}{c|}{F1}     & DR   &-  \\ \hline \hline
GPT-2                  & \multicolumn{1}{c|}{0.8736} & \multicolumn{1}{c|}{0.2982} & 0.3661 & \multicolumn{1}{c|}{0.5130} & \multicolumn{1}{c|}{0.0326} & 0.0563 & \multicolumn{1}{c|}{0.5099} & \multicolumn{1}{c|}{0.0279} & 0.0380 & \multicolumn{1}{c|}{0.9681} & \multicolumn{1}{c|}{0.7887} & 0.8158 & 0.4407 \\ \hline
NetGPT                 & \multicolumn{1}{c|}{0.9087} & \multicolumn{1}{c|}{0.4627} & 0.5241 & \multicolumn{1}{c|}{0.5390} & \multicolumn{1}{c|}{0.0526} & 0.0870 & \multicolumn{1}{c|}{0.5124} & \multicolumn{1}{c|}{0.0299} & 0.0418 & \multicolumn{1}{c|}{0.9692} & \multicolumn{1}{c|}{0.8231} & 0.8421 & 0.4827\\ \hline
NetGPT-A               & \multicolumn{1}{c|}{0.8878} & \multicolumn{1}{c|}{0.3554} & 0.4243 & \multicolumn{1}{c|}{0.5530} & \multicolumn{1}{c|}{0.0652} & 0.1023 & \multicolumn{1}{c|}{0.5160} & \multicolumn{1}{c|}{0.0363} & 0.0461 & \multicolumn{1}{c|}{0.9681} & \multicolumn{1}{c|}{0.8130} & 0.8421 & 0.4675\\ \bottomrule
\end{tabular}}
\end{table*}
\begin{figure*}
\begin{minipage}{.47\linewidth}
\centering
\includegraphics[width=0.95\textwidth]{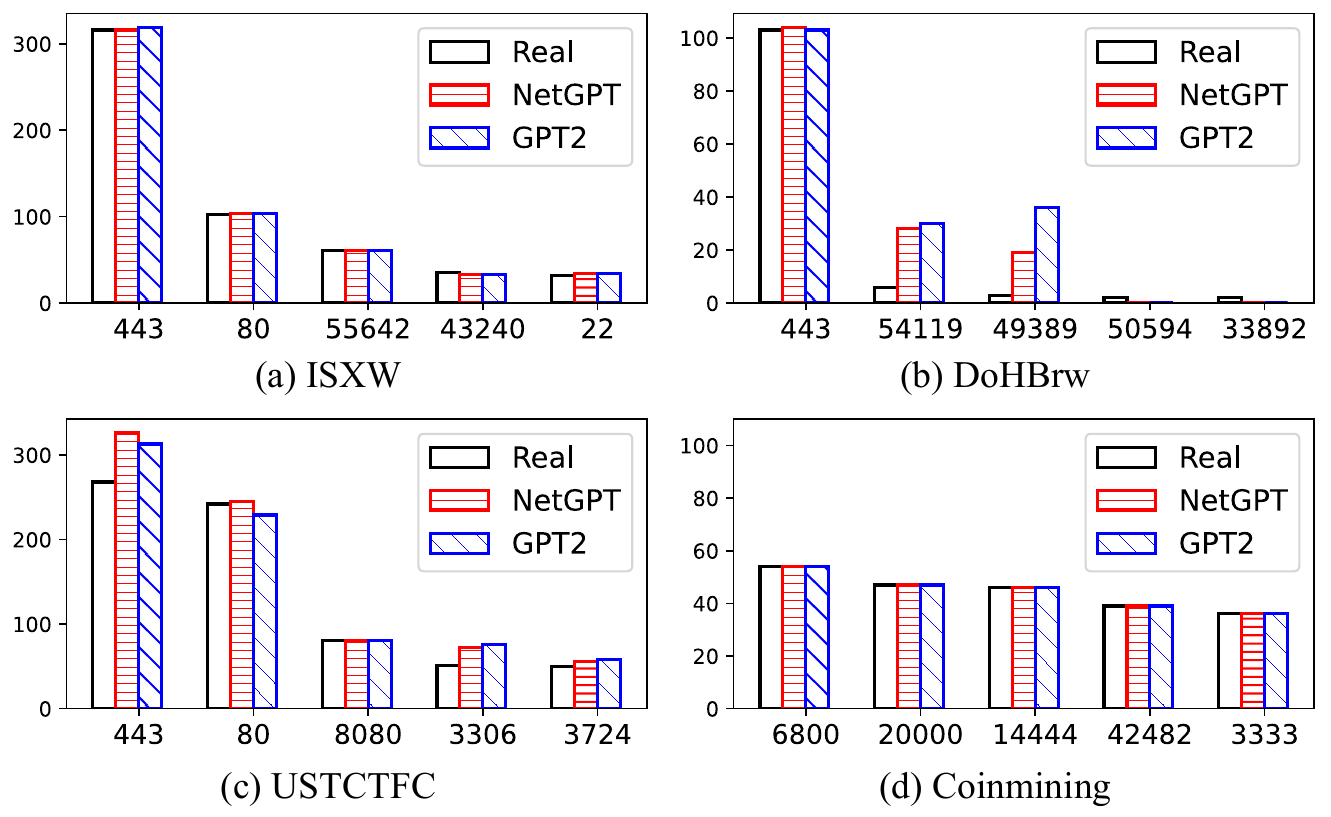}
\caption{Top 5 source ports.}
\label{fig:sport}
\end{minipage}
\begin{minipage}{.47\linewidth}
\centering
\includegraphics[width=0.95\textwidth]{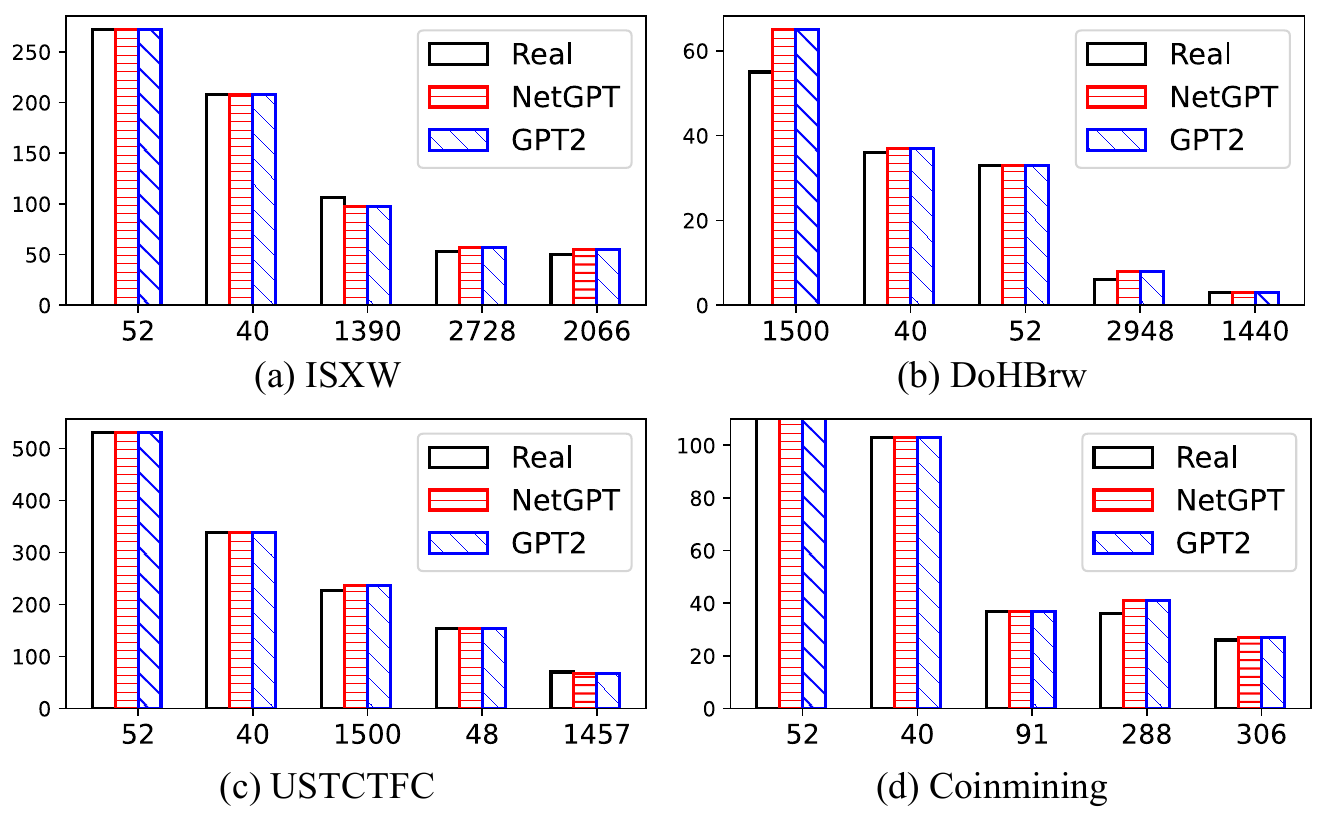}
\caption{Top 5 packet length.}
\label{fig:len}
\end{minipage}
\end{figure*}
\section{Appendix}

\subsection{Additional Metric Results on Traffic Generation}
We also evaluate the generation performance by Accuracy (AC), Macro F1 (F1) and Diversity Ratio (DR). The DR is calculated by dividing the number of unique generated texts by the total number of possible generated texts. NetGPT-A is finetuned by all targeted header fields and all datasets simultaneously.
 
From Table \ref{tb:TG1}, we can make the following observations.
\textbf{(1)} NetGPT achieves the best average performance and outperforms GPT-2 in all the datasets on all metrics, demonstrating the effectiveness of shuffling in the traffic generation tasks. \textbf{(2)} NetGPT-A receives lower average performance than NetGPT, however, outperforms NetGPT in two datasets, which may be because similar traffic may have different network header fields in different datasets. \textbf{(3)} NetGPT can achieve much better DR which may help broaden the testing scopes for network behavior simulation in the real world.

\subsection{Additional Visual Results on Traffic Generation}
The visual distributions of top-K source ports and length are presented in Figure \ref{fig:sport} and \ref{fig:len}. Compared to direct GPT-2, NetGPT can generate relatively similar distributions.

\end{document}

%% file: intro.tex
\section{Introduction} 
\begin{figure}
\centering
\includegraphics[width=0.5\textwidth]{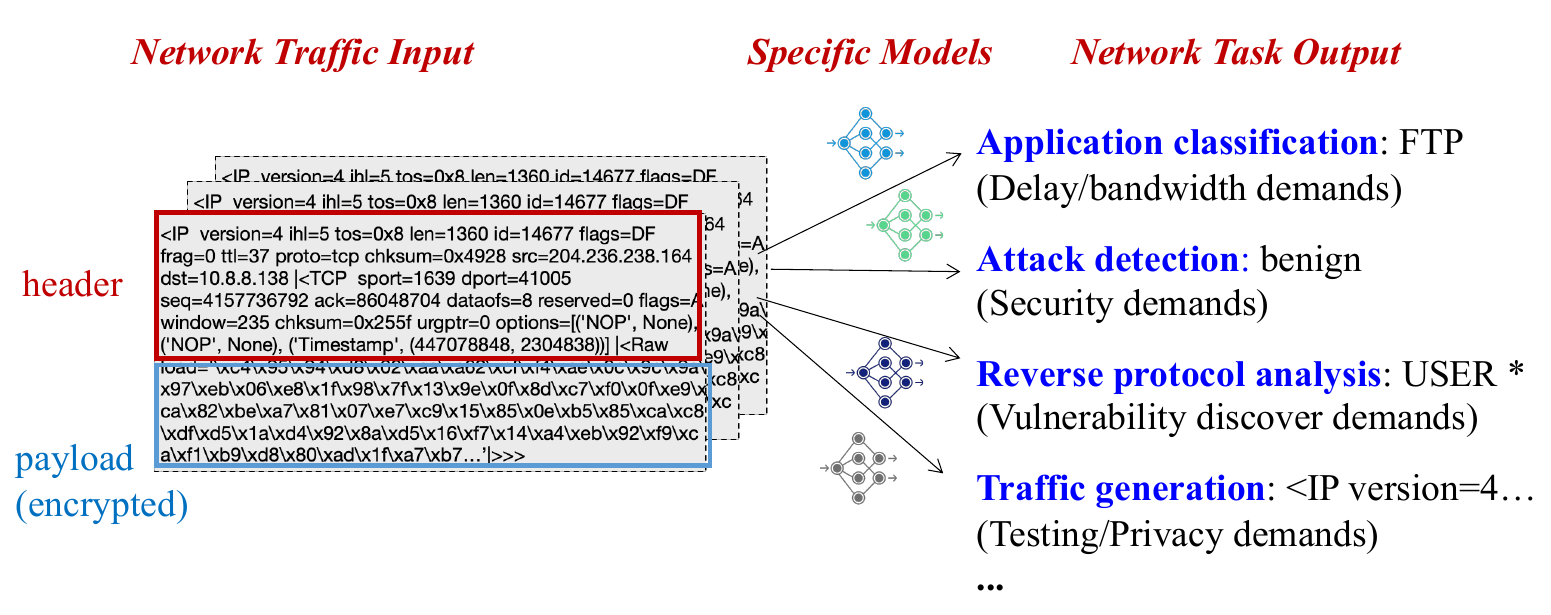}
\vskip -1em
\caption{An example of diverse downstream traffic tasks.}
\vskip -2em
\label{fig:pretrain}
\end{figure}

All data on the Internet are transferred by network traffic, which in turn can reveal network service quality, attack conditions, user network behavior patterns, etc. To provide high-quality network services as well as ensure data privacy, there are plenty of network tasks for diverse network service demands. With the development of artificial intelligence, recent years have witnessed great advancements in intelligent networks to improve these task performances from both academia and industry. As shown in Figure \ref{fig:pretrain}, to satisfy demands of building a better network environment, existing works design specific models for different network tasks, including encrypted application classification \cite{LinXGLSY22}, attack detection \cite{MengWWYZ21},  reverse protocol analysis \cite{ZhangMZ22}, etc. Meanwhile, to ensure data privacy, traffic generation and trace synthesizing are carefully studied \cite{Yin0JFS22}. However, they always customize specific models for specific network tasks, which may suffer limitations from small sample sizes for specific tasks, insufficient model training under small samples, and high costs to develop customized models. Note that, we use traffic to uniformly denote both packets and flows.

To relieve limitations of customization, a pretraining strategy is proposed to utilize large-scale data to learn intrinsic characteristics without considering specific downstream tasks \cite{bert,gpt2}. The pretrained model can easily adapt to diverse downstream tasks, and help optimize the efficiency and efficacy, which has recently received great success in natural language processing \cite{gpt3,gpt3.5}. Motivated by these works, we assume the pretrained model of network traffic can help learn the intrinsic characteristics of network traffic and adapt to downstream network tasks. Previous work has demonstrated that pretrained representation models can significantly enhance various network tasks such as application classification and attack detection \cite{MengWMLLZ22}. However, although existing works do map network traffic into latent semantic space, they still target specific tasks and can not adapt to other tasks, especially for traffic generation tasks \cite{LinXGLSY22,MengWMLLZ22}. 
Considering the specialty of raw traffic input and network task output, existing generative pretrained models for NLP cannot be directly applied in the network field. Building a generative pretrained model for network traffic faces multiple challenges.

First, for raw traffic inputs in the pretraining process, considering the heterogeneous headers and payloads in multi-pattern network traffic, it is difficult to integrate semantic information effectively. 
On the one hand, with the continuous enrichment of network services, network protocol types are also constantly increasing, especially in the industrial Internet where there are a large number of private protocols \cite{ZhangMZ22}. The syntax structure of different network traffic's header and payload parts has significant differences. On the other hand, except for plaintext, various encrypted traffic exists in the network. The division of tokens and construction of vocabularies face challenges and it is difficult to retain rich semantics of both plaintext and ciphertext for both diverse network tasks.

Second, for network task outputs in the finetuning process, different network tasks have different dependencies on contexts. 
On the one hand, traffic understanding tasks, such as application classification and attack detection, require to integrate bidirectional information (like BERT \cite{bert}). However, traffic generation tasks, such as testing traffic generation and network behavior simulation, only need to use already generated information (like GPT-based methods \cite{gpt3,gpt3.5,gpt2}) to learn unidirectional information, whereas resulting in reduced effectiveness for traffic understanding. 
Since traffic generation is a core issue for related tasks, we will follow the unidirectional way to support both traffic understanding and generation tasks, which makes it necessary to handle the cost of relatively insufficient modeling of token associations. On the other hand, to adapt to diverse tasks including both packet-level and flow-level ones, it is important to  seize the sequential contexts of packets as well as retain the packet structure.
However, the packet count and length (Ethernet MTU is 1500) can be relatively large, and as packet structure reflects protocol standards, tokens inside a packet should not be split. It is still challenging to adapt to diverse tasks and optimize downstream performance.

To tackle these challenges, we propose NetGPT to learn essential characteristics of traffic and effectively improve the performance of diverse network tasks. To the best of our knowledge, this is the \textit{first} attempt to propose a pretrained model for traffic understanding and generation tasks. In the pretraining process, we construct general text inputs based on the hex, which can maintain the semantics of both plaintext and ciphertext, and can also be easily transferred to practicable plaintext traffic in the traffic generation. Based on this general encoding, we encode multi-pattern network traffic into a general semantic space and obtain a foundational pretrained model that supports diverse tasks. In the finetuning process, based on the observation of independent network header fields, we enlarge the usable information and learn from the latter information by shuffling header fields, segmenting packet semantics during flow-level training, and incorporating labels of diverse traffic understanding and generation tasks with designed prompts. In this way, we optimize the pretrained model of network traffic for various tasks, supporting both traffic understanding and generation tasks with great performance. 
With traffic datasets from encrypted software, DNS, private industrial protocols and cryptocurrency mining, expensive experiments demonstrate the effectiveness of our NetGPT in a range of traffic understanding and generation tasks on traffic datasets, and outperform state-of-the-art baselines by a wide margin.

%% file: related.tex
\begin{figure*}[]
\centering
\includegraphics[width=0.99\textwidth]{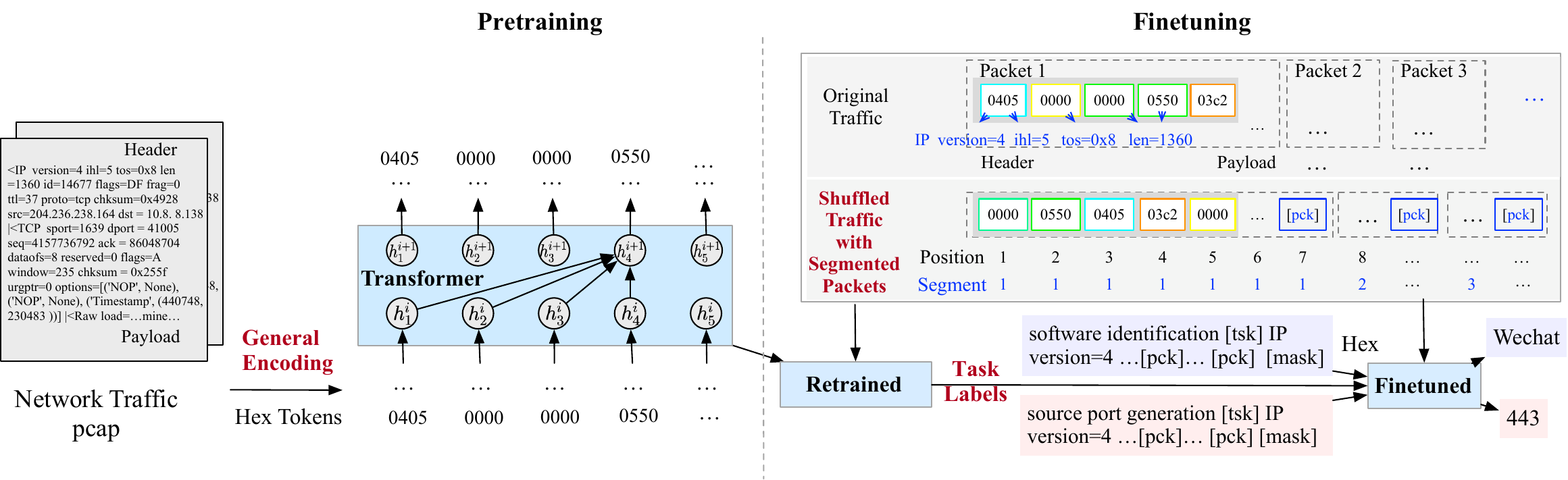}
\caption{The framework of \modelname.}
\label{fig:framework}
\end{figure*}

\section{Related Work} 
\subsection{Intelligent Network Tasks}
Intelligent network tasks can be classified into two categories based on their nature, i.e., traffic understanding tasks and traffic generation tasks.

The goal of traffic understanding tasks is to identify and classify network traffic for the purpose of management, security, and performance optimization \cite{DainottiPC12}. Existing research on traffic understanding tasks can be classified into feature-based and byte-based ones. Feature-based methods construct effective features tailored to specific applications from raw traffic data to improve the model's performance on specific applications \cite{MengWWYZ21,ShenZZ0DL19,ZhengGYM20}. Byte-based methods automatically mine deep features from raw traffic data to enable the model to more accurately capture the distribution of the original traffic data \cite{CasinoCP19,MinLLCC18,ZhangLYW20}. However, due to diverse data representation patterns, these methods have difficulty extracting effective information at the semantic and syntactic levels, resulting in poor model performance and a lack of generality for different tasks.

Traffic generation tasks create or simulate network traffic for performance evaluation and benchmark testing. Existing research on traffic generation tasks primarily revolves around Generative Adversarial Networks (GANs) \cite{HuiWWYLJL22,abs-2107-14776,RingSLH19,Yin0JFS22}. Ring et al. \cite{RingSLH19} first propose using GANs to generate flow-level test traffic. Yin et al. \cite{Yin0JFS22} propose ensuring fidelity while generating test traffic using GANs. However, GAN-based approaches for generating network traffic are limited by their own models as they lack generality. Additionally, current generated flows are mainly tabular data which loses the information of packet structure, semantics and sequences.

In summary, current intelligent network tasks still rely on constructing specific models to improve model performance for specific tasks, and these models cannot be used for other network tasks.

\subsection{Traffic Representation Learning}
Traffic representation learning is an important foundation for network security and management. Similar to pretraining, it transforms raw traffic data into a low-dimensional space that accentuates traffic patterns, characterizes traffic attributes, and unveils the similarities and dissimilarities among various types of traffic \cite{LinXGLSY22,PapadogiannakiI21}.
It serves as an intermediary step in downstream endeavors, such as attack detection and application classification. There are two broad categories of traffic representation learning, namely feature-based and byte-based.

Feature-based traffic representation learning models extract representative features of various attacks as input for the model, and then learn to train low-dimensional representations with discrimination using feature vectors and classification labels \cite{HypoliteSHDDS20,ShenZZ0DL19}. These approaches serve as an intermediate step for tasks related to network traffic understanding. However, due to the varying requirements for extracting network traffic features in different application scenarios, current feature-based approaches are difficult to apply simultaneously across multiple network applications.

Byte-based traffic representation learning models do not require complex feature extraction or processing. Instead, they directly use byte information from the original packet data (i.e., representing feature vectors as hex numbers), which are then mapped into a low-dimensional space that distinguishes different labels \cite{CasinoCP19,JemalHCM21,RingSLH19,ZhangLYW20}. In traffic understanding tasks, byte information is generally treated as pixel values in images \cite{JemalHCM21,LinXGLSY22,MengWMLLZ22}, but this approach wastes rich semantic information between bytes. For traffic generation tasks, the mainstream approach is to treat byte information as tabular data \cite{RingSLH19}. However, these approaches cannot capture packet count or length information adequately nor can they fully exploit inter-packet dependencies within a single flow.

In conclusion, although traffic representation can transform traffic into semantic space like pretraining work does, current research on traffic representation learning mainly focuses on designing specific models for particular application scenarios. Application-specific models tend to lose original flow information, and cannot be applied simultaneously across multiple application scenarios.

%% file: core.tex
\section{\modelname} 
In this section, we will present our proposed \modelname, in order to provide a general model for network traffic in different patterns and easily adapt to both traffic understanding tasks (e.g., attack detection, application classification) and traffic generation tasks (e.g., network equipment testing, synthetic trace generation). Note that, as one flow can be made up of only one packet, we use a general term, namely network traffic, to denote both flows and packets.

As shown in Figure \ref{fig:framework}, we have two important processes, i.e., the pretraining and the finetuning.
In the pretraining process, we train a pretrained model to encode multi-pattern network traffic into a general semantic space. In the finetuning process, we optimize the corresponding inference performance with pretrained model and specific task labels.

\subsection{Pretraining}
\label{sec:pre}
Considering different patterns of network traffic, we first introduce the general encoding strategy to encode multi-pattern network traffic into a general semantic space. Furthermore, to maintain the generalization ability across different tasks, we utilize datasets collected from different network scenarios without any labels in the pretraining process.

\textbf{General Encoding.}
Network traffic exhibits various patterns due to several factors. Firstly, the presence of diverse protocols in different layers of protocol stacks results in varying formats of network traffic. Secondly, even within the same protocol, network traffic can differ across the specific network services. For instance, different websites may generate varying traffic sizes and patterns. Finally, to safeguard network security and prevent unauthorized inspection of data, encryption ciphers are often implemented through the use of network security protocols. As a result, the original patterns of network traffic may be obscured. 

Furthermore, except for the various traffic patterns, the encoding requirements and downstream tasks differ when dealing with a composite of plaintext and encrypted network traffic. Specifically, in the case of encrypted traffic, the model can only handle traffic understanding tasks. Conversely, in plaintext traffic, the model must be adept in both traffic understanding and plaintext generation. 

To learn various traffic patterns as well as handle the different encoding requirements, we translate each byte into its corresponding hex number, and then use a tokenizer to generate tokens. In this way, different kinds of traffic patterns can be learned from the uniformed hex patterns, and hex numbers can be easily transferred back to plaintext with UTF-8.
Specially, as each byte can only represent up to 255 kinds of tokens in the vocabulary space (i.e., the maximal hex number is ``ff''), it may be not sufficient for capturing the full extent of underlying semantic information. Therefore, we reconstruct a more comprehensive vocabulary that can accommodate more complex semantic information with Wordpiece~\cite{bert}.

\textbf{Objective Function.}
To ensure the generalization ability across different tasks, we employ the auto-regressive model GPT-2 as the base model for our pretraining process. This model predicts the next token based on the previous token(s), and is therefore well-suited for our understanding and generation demands. Specifically, given an input traffic sequence consisting of L tokens, denoted as $\mathbf{x}=\{t_{1},t_{2},\dots,t_{L}\}$, our \modelname calculates the probability $P_L$ of token $t_k$ based on the preceding $k-1$ tokens:
\begin{equation}
    P_L(t_k | t_1, \dots, t_{k-1}) = \softmax(W_vh_{{k-1}})
\end{equation}
Here, $h_{{k-1}}$ denotes the representation encoded by Transformer with the previous tokens $\{t_1, \ldots, t_{k-1}\}$ as input. $W_{v}$ represents the learnable parameters. With these definitions, we have the objective function.
\begin{equation}
	\mathcal{J}(\theta) = \sum\sum_k \log P_L(t_k | t_1, \dots, t_{k-1}) \label{eq:loss1}
\end{equation}
In this paper, we employ network traffic to denote both packets and flows. However, it is observed that the number of packets within each flow can vary significantly~\cite{FuLSX21}. Considering the maximum transmission unit of each packet is up to 1500 bytes, and processing an excessive amount of traffic data can lead to unnecessary computational overhead. To address these issues, as some works demonstrate the first packets of flows are more important~\cite{DaiXX23}, we only include the first three packets of heavy flows.

\subsection{Finetuning} 
With the pretrained model and specific task labels, we take full advantage of the characteristics of network traffic and optimize the corresponding inference performance. Specifically, we enlarge the usable information and learn from the latter information by shuffling header fields, segmenting packet semantics during flow-level training, and incorporating labels of diverse traffic understanding and generation tasks with designed prompts.

\textbf{Shuffling Header Fields.}
To cater to adapt to generation tasks, we use GPT models as our basic model.
However, this comes at a cost of diminished understanding performance. Unlike BERT-based models that can utilize bi-direction information, GPT-based models learn the representation of $k$-th token $t_k$ solely based on the preceding $k-1$ tokens. Hence, they can only access one-directional information. The valuable information that is present after $t_k$ cannot be incorporated in, thereby negatively impacting downstream performance.

Unlike conventional pretraining tasks in natural language processing, the header fields of network traffic are relatively independent. 
This unique characteristic can be leveraged to significantly enhance the performance of one-directional learning. 
Based on that, we shuffle among header fields and optimize both the pretrained model and downstream tasks without affecting the semantics of the corresponding traffic. Also, this shuffling can produce more data which can alleviate the small sample size problem.
Note that although the changing orders of header fields have no impact on semantics, there are still a few tasks (such as generating testing traffic for private protocols) that require the correct header field order. Therefore, we do not conduct this shuffling procedure during pretraining but instead in the finetuning process.

In detail, as each header field contains complete semantic information and our general encoding is based on bytes, shuffling should only be performed among network header fields with a minimum shuffle unit of one byte.
For instance, based on the IP protocol formats (as illustrated in Figure \ref{fig:framework}), the \textit{version} (ip version) and \textit{header length} fields (ihl) occupy the first byte, the \textit{service type} field (tos) occupies the second byte, and the \textit{total length} field (len) spans across the third and fourth bytes. In this example, the \textit{version} and \textit{header length} fields should be treated as a single unit for shuffling purposes, while the third and fourth bytes should also be considered as a single unit for shuffling.

\textbf{Segmenting Packets in Flows.}
Packets belonging to the same flow share identical five-tuple header fields, which include source IP, source port, destination IP, destination port and protocol. However, these packets may not exhibit semantic similarities. 
Even in flow-level tasks, the content and order of packets within a single flow are of great importance. For example, packet length sequence in a flow is one of the most important features for flow-level encrypted traffic classification tasks \cite{FuLSX21}.
Therefore, it is essential to effectively segment packets within a flow. Additionally, our experimental results illustrate that treating a flow as a whole and ignoring the characteristics of each packet can negatively impact downstream flow-level performance (as shown in Section \ref{sec:exp}). Furthermore, since our \modelname aims to maintain generalization ability for both packets and flows, we only segment packets during finetuning for flow-level tasks.

To preserve the semantics and sequence of packets, as depicted in Figure \ref{fig:framework}, we first concatenate packets sharing the same five-tuple based on their timestamps to form a flow. Inspired by the next sentence prediction approach of BERT \cite{bert}, we append a special character [pck] to the end of each packet as a delimiter and incorporate segment embeddings as an indicator of each packet.
To adhere to the standard encoding process outlined in Section \ref{sec:pre}, this special character is also encoded into hex.
Thus, for each flow, we can have the classical delimiter [cls] to represent the representation of the entire flow. This representation can be utilized as input for downstream tasks such as flow classification. 
Similarly, since our one-direction model progresses from left to right, [pck] can serve as the representation results for each packet. In this way, we can preserve both semantic and sequential information for each packet in a flow simultaneously. 

\begin{figure}
\centering
\includegraphics[width=0.45\textwidth]{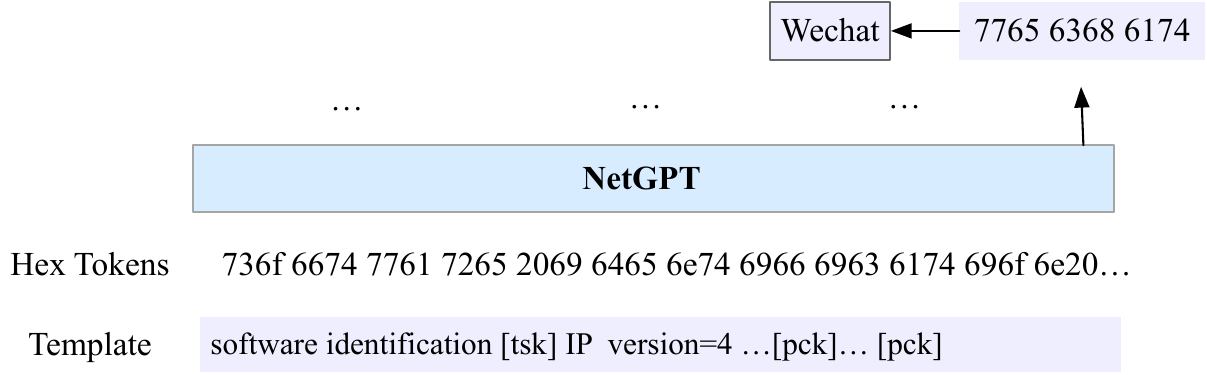}
\caption{Inference example with prompts.}
\label{fig:prompt}
\vskip -1em
\end{figure}

\textbf{Incorporating Diverse Task Labels.}
By employing header field shuffling and packet segmenting modules, we optimize the input traffic for fine-tuning. These optimized inputs can be further enhanced through the use of labels and guidance, which are typically applied in two ways for understanding and generation tasks. For understanding tasks, a linear classifier is trained based on the representation of the [cls] token (called the [cls] way). For generation tasks, the targeted tokens are generated based on the last token of the input (called the text2text way). However, with these two ways, there are two limitations. 

On one hand, these two ways rely solely on the representation of a single token (i.e., [cls] token and the last token), thereby rendering field shuffling ineffective.
To tackle this limitation, we include a quick retraining with shuffled data from the corresponding dataset (as illustrated in Figure \ref{fig:framework}), based on Eq.(\ref{eq:loss1}). Note that, to keep consistency with the vocabulary of the pretraining process, we do not use special tokens or packet segmenting during the retraining process. 

On the other hand, the model is trained by the text2text way in the pretraining process, thus there are inconsistencies between pretraining and finetuning when using the [cls] way. To mitigate this issue, we reformulate traffic understanding tasks as generation tasks with different prompts \cite{gpt3}. Specially, as shown in Figure \ref{fig:prompt}, to adapt to diverse network tasks, we train different kinds of tasks simultaneously. For traffic understanding tasks, we utilize the task name as a prompt, translate it into hex, attach it to the beginning of inputs, and try to generate the right outputs. Additionally, for traffic generation tasks, to generate different kinds of network headers, we use the header field as hex prompts to generate targeted tokens with a single model for each dataset. 
The position and segment embeddings will also be updated for the prompts accordingly.

%% file: experiment.tex
\section{Experiments} 
\label{sec:exp}

\subsection{Settings} 

\begin{table}[]
\centering
\caption{Statistics of the datasets.}
\scalebox{0.75}{
\begin{tabular}{l|c|c|c|c|c}
\hline
\textbf{Dataset}                                                  & ISXW & DoHBrw & USTCTFC & PrivII & Cybermine\\ \hline\hline
\#  packets                                       & 1,338,300      & 77,149,018  &  1,010,534 & 38,250,867 & 7,862 \\ \hline
\#  flows                                       & 1,262      & 2,497  &  5,039 & 36,926 &  2,617\\ \hline
\end{tabular}}
\label{tb:stat}
\end{table}

\begin{table*}[]
\centering
\caption{Traffic understanding performance comparison.}
\label{tb:TU}
\scalebox{0.75}{
\begin{tabular}{c|cc|cc|cc|cc|cc|cc|cc|cc|c}
\toprule
\multirow{2}{*}{model}                                      & \multicolumn{2}{c|}{task 1}          & \multicolumn{2}{c|}{task 2}          & \multicolumn{2}{c|}{task 3}          & \multicolumn{2}{c|}{task 4}          & \multicolumn{2}{c|}{task 5}          & \multicolumn{2}{c|}{task 6}          & \multicolumn{2}{c|}{task 7}          & \multicolumn{2}{c|}{task 8}          & Avg\\ \cline{2-18} 
                                                            & \multicolumn{1}{c|}{AC}     & F1     & \multicolumn{1}{c|}{AC}     & F1     & \multicolumn{1}{c|}{AC}     & F1     & \multicolumn{1}{c|}{AC}     & F1     & \multicolumn{1}{c|}{AC}     & F1     & \multicolumn{1}{c|}{AC}     & F1     & \multicolumn{1}{c|}{AC}     & F1     & \multicolumn{1}{c|}{AC}     & F1    & - \\ \hline\hline
\begin{tabular}[c]{@{}c@{}}ET-BERT\\ (packet)\end{tabular}  & \multicolumn{1}{c|}{0.9990}  & 0.9990  & \multicolumn{1}{c|}{0.9980}  & 0.9970  & \multicolumn{1}{c|}{1.0000} & 1.0000 & \multicolumn{1}{c|}{0.9624} & 0.9624 & \multicolumn{1}{c|}{1.0000} & 1.0000 & \multicolumn{1}{c|}{0.9636} & 0.8664 & \multicolumn{1}{c|}{1.0000} & 1.0000 & \multicolumn{1}{c|}{0.9952} & 0.9771 & 0.9825\\ \hline
\rowcolor[HTML]{EFEFEF}\begin{tabular}[c]{@{}c@{}}ET-BERT\\ (flow)\end{tabular}    & \multicolumn{1}{c|}{0.9375} & 0.9372 & \multicolumn{1}{c|}{0.9206} & 0.4314 & \multicolumn{1}{c|}{1.0000} & 1.0000 & \multicolumn{1}{c|}{0.8320}  & 0.8187 & \multicolumn{1}{c|}{1.0000} & 1.0000 & \multicolumn{1}{c|}{0.9524} & 0.6986 & \multicolumn{1}{c|}{1.0000} & 1.0000 & \multicolumn{1}{c|}{1.0000} & 1.0000 & 0.9080\\ \hline
\begin{tabular}[c]{@{}c@{}}GPT-2\\ (packet)\end{tabular}    & \multicolumn{1}{c|}{1.0000} & 1.0000 & \multicolumn{1}{c|}{0.9964} & 0.9952 & \multicolumn{1}{c|}{1.0000} & 1.0000 & \multicolumn{1}{c|}{0.9132} & 0.9130  & \multicolumn{1}{c|}{1.0000} & 1.0000 & \multicolumn{1}{c|}{0.9499} & 0.9481 & \multicolumn{1}{c|}{1.0000} & 1.0000 & \multicolumn{1}{c|}{1.0000} & 1.0000 & 0.9822\\ \hline
\rowcolor[HTML]{EFEFEF}\begin{tabular}[c]{@{}c@{}}GPT-2\\ (flow)\end{tabular}      & \multicolumn{1}{c|}{0.8750}  & 0.8667 & \multicolumn{1}{c|}{0.9286} & 0.7356 & \multicolumn{1}{c|}{1.0000} & 1.0000 & \multicolumn{1}{c|}{0.8240}  & 0.8001 & \multicolumn{1}{c|}{1.0000} & 1.0000 & \multicolumn{1}{c|}{0.9563} & 0.9460  & \multicolumn{1}{c|}{1.0000} & 1.0000 & \multicolumn{1}{c|}{1.0000} & 1.0000 & 0.9333\\ \hline\hline
\begin{tabular}[c]{@{}c@{}}\textbf{NetGPT}\\ (packet)\end{tabular}   & \multicolumn{1}{c|}{1.0000} & 1.0000 & \multicolumn{1}{c|}{0.9974} & 0.9970  & \multicolumn{1}{c|}{1.0000} & 1.0000 & \multicolumn{1}{c|}{0.9308} & 0.9301 & \multicolumn{1}{c|}{1.0000} & 1.0000 & \multicolumn{1}{c|}{0.9575} & 0.9567 & \multicolumn{1}{c|}{1.0000} & 1.0000 & \multicolumn{1}{c|}{1.0000} & 1.0000 & 0.9856\\ \hline
\rowcolor[HTML]{EFEFEF}\begin{tabular}[c]{@{}c@{}}\textbf{NetGPT}\\ (flow)\end{tabular}     & \multicolumn{1}{c|}{0.9375} & 0.9352 & \multicolumn{1}{c|}{0.9683} & 0.8056 & \multicolumn{1}{c|}{1.0000} & 1.0000 & \multicolumn{1}{c|}{0.8120}  & 0.7747 & \multicolumn{1}{c|}{1.0000} & 1.0000 & \multicolumn{1}{c|}{0.9563} & 0.9463 & \multicolumn{1}{c|}{1.0000} & 1.0000 & \multicolumn{1}{c|}{1.0000} & 1.0000 & 0.9460\\ \hline
\begin{tabular}[c]{@{}c@{}}\textbf{NetGPT-A}\\ (packet)\end{tabular} & \multicolumn{1}{c|}{1.0000} & 1.0000 & \multicolumn{1}{c|}{0.9966} & 0.9955 & \multicolumn{1}{c|}{1.0000} & 1.0000 & \multicolumn{1}{c|}{0.9332} & 0.9332 & \multicolumn{1}{c|}{1.0000} & 1.0000 & \multicolumn{1}{c|}{0.9563} & 0.9555 & \multicolumn{1}{c|}{1.0000} & 1.0000 & \multicolumn{1}{c|}{1.0000} & 1.0000 & 0.9856\\ \hline
\rowcolor[HTML]{EFEFEF}\begin{tabular}[c]{@{}c@{}}\textbf{NetGPT-A}\\ (flow)\end{tabular}   & \multicolumn{1}{c|}{0.9375} & 0.9352 & \multicolumn{1}{c|}{0.9286} & 0.7220 & \multicolumn{1}{c|}{1.0000} & 1.0000 & \multicolumn{1}{c|}{0.8000}  & 0.7743 & \multicolumn{1}{c|}{1.0000} & 1.0000 & \multicolumn{1}{c|}{0.9563} & 0.9456 & \multicolumn{1}{c|}{1.0000} & 1.0000 & \multicolumn{1}{c|}{1.0000} & 1.0000 & 0.9375\\ \bottomrule
\end{tabular}}
\end{table*}
\textbf{Datasets.}
\label{sec:data}
We collect five datasets comprising a total of 113GB in size, including three publicly available datasets (ISXW 2016\footnote{https://www.unb.ca/cic/datasets/vpn.html}, DoHBrw 2020\footnote{https://www.unb.ca/cic/datasets/dohbrw-2020.html} and USTCTFC 2016\footnote{https://github.com/yungshenglu/USTC-TFC2016}), one self-simulated traffic dataset PrivII 2021 that mimics private protocols used in the industrial internet, and one self-constructed cryptojacking dataset called Cybermining 2023. The statistics of datasets used in our experiments are summarized in Table \ref{tb:stat}.
Note that the Cybermining dataset is not used during the pretraining phase and will be utilized as an unseen dataset for assessing generalization.

Among these datasets with labels, for the \textit{traffic understanding tasks}, we refer to prior research \cite{MengWMLLZ22} to formulate eight tasks. In detail, in ISXW, we have VPN detection (\textit{task 1}) with two classes and application classification (\textit{task 2}) with 13 classes; in DoHBrw, we have malicious DoH query detection (\textit{task 3}) with two classes and DoH query generator identification (\textit{task 4}) with 5 classes; in USTCTFC, we have attack detection (\textit{task 5}) with 2 classes and software identification (\textit{task 6}) with 20 classes; in Cybermining, we have cybermining detection (\textit{task 7}) with 2 classes and cryptocurrency identification (\textit{task 8}) with 7 classes. 

In addition, for \textit{traffic generation tasks}, we generate five important header fields based on \cite{Yin0JFS22}, namely source IP, destination IP, source port, destination port and length. With these important synthetic header fields, it is easy to construct usable pcap traffic to benchmark and test new hardware and software capabilities by concatenated and stuffed bytes for payloads \cite{Yin0JFS22,ZhangMZ22}.
Each dataset is trained separately in a self-supervised finetuning process.

\textbf{Baselines.} 
We compare with two most representative pretraining works, namely a BERT-based one (ET-BERT \cite{LinXGLSY22}) and a GPT-based one (GPT-2 \cite{gpt2}).
As ET-BERT has demonstrated superior performance in traffic classification tasks compared to the basic BERT \cite{bert} and all other non-pretraining methods, we opt to use it instead of BERT and will not include the traditional non-pretraining ones. Furthermore, given that our inputs differ from those of natural languages, we have to retrain the pretraining process and cannot utilize newer GPT-based models such as GPT-3 \cite{gpt3} and GPT-3.5 \cite{gpt3.5} which only provide pretrained language models. We take the GPT-2-base with 0.1B parameters as our baseline and basic model.

To provide fair comparisons, we train the NetGPT and baselines with the same inputs by general encoding in the pretraining process. In the finetuning process, we conduct traffic understanding and generation tasks in different ways.
For the \textit{traffic understanding tasks}, we compare our NetGPT with both baselines in both packet-level and flow-level ways. For the \textit{traffic generation tasks}, we solely compare with GPT-2 in a packet-level way, since ET-BERT does not possess any generation capabilities and header fields for packets in a flow differ from one another.

\textbf{Metrics and Implementation Details.} We evaluate the performance by Accuracy (AC), Macro F1 (F1) and Jensen-Shannon Divergence (JSD). The AC and F1 are widely used for traffic understanding tasks, e.g., application classification and attack detection \cite{LinXGLSY22,MengWMLLZ22}. The JSD can measure the fidelity in traffic generation tasks based on the comparison between the real and generated data for domain-relevant distribution \cite{xu2021stan,Yin0JFS22}. Note that higher AC and F1 represent better performance, while lower JSD denotes better fidelity.

We conduct the experiments with a random selection of 5000 packets for validation and another 5000 for testing. The remaining packets are used for pretraining. In the pretraining process, we remove headers to avoid biased interference based on the settings of ET-BERT \cite{LinXGLSY22}. With a maximum token size of 512, we only use the first three packets if the flow is very long. We set the batch size to 96, total steps to 500,000, learning rate to $2\times10^{-5}$, and warmup ratio to 0.1. In the finetuning process, for both traffic generation and traffic understanding tasks, we set the batch size to 32 and input sequence length to 256 and the target output sequence length is set to 4. Specially, for the Shuffling Header Fields module, we randomly select two header fields of the packet, and exchange them to generate new data. For traffic understanding tasks, we fine-tune with AdamW optimizer for 50 epochs while for traffic generation tasks, we fine-tune with AdamW optimizer for only 10 epochs. All experiments are conducted on V100-32GB GPUs.

\begin{table*}
\centering
\caption{Traffic generation performance comparison on JSD.}
\label{tb:TG}
\scalebox{0.75}{
\begin{tabular}{c|ccc|ccc|ccc|ccc|c}
\toprule
 \multirow{2}{*}{model} & \multicolumn{3}{c|}{ISXW}                                          & \multicolumn{3}{c|}{DoHBrw}                                        & \multicolumn{3}{c|}{USTCTFC}                                       & \multicolumn{3}{c|}{Cybermining}  & Avg                                 \\ \cline{2-14} 
                       & \multicolumn{1}{c|}{len}     & \multicolumn{1}{c|}{dport}     & sport     & \multicolumn{1}{c|}{len}     & \multicolumn{1}{c|}{dport}     & sport     & \multicolumn{1}{c|}{len}     & \multicolumn{1}{c|}{dport}     & len     & \multicolumn{1}{c|}{dport}     & \multicolumn{1}{c|}{sport}     & len   &-  \\ \hline \hline
GPT-2                 & \multicolumn{1}{c|}{0.0844} & \multicolumn{1}{c|}{0.0062} & 0.0133 & \multicolumn{1}{c|}{0.0611} & \multicolumn{1}{c|}{0.0044} & 0.0067 & \multicolumn{1}{c|}{0.0193} & \multicolumn{1}{c|}{0.1917} & 0.1105 & \multicolumn{1}{c|}{0.0027} & \multicolumn{1}{c|}{0.0000} & 0.0000 & 0.0417\\   \hline
\textbf{NetGPT}                  & \multicolumn{1}{c|}{0.0492} & \multicolumn{1}{c|}{0.0029} & 0.0074 & \multicolumn{1}{c|}{0.0596} & \multicolumn{1}{c|}{0.0039} & 0.0115 & \multicolumn{1}{c|}{0.0186} & \multicolumn{1}{c|}{0.2093} & 0.1222 & \multicolumn{1}{c|}{0.0027} & \multicolumn{1}{c|}{0.0000} & 0.0000 & 0.0406 \\\bottomrule
\end{tabular}}
\end{table*}

\begin{figure*}
	\centering
	\subfigure[CDF on ISXW2016]{
	\begin{minipage}[t]{0.49\linewidth}
	\centering
\includegraphics[width=1\textwidth]{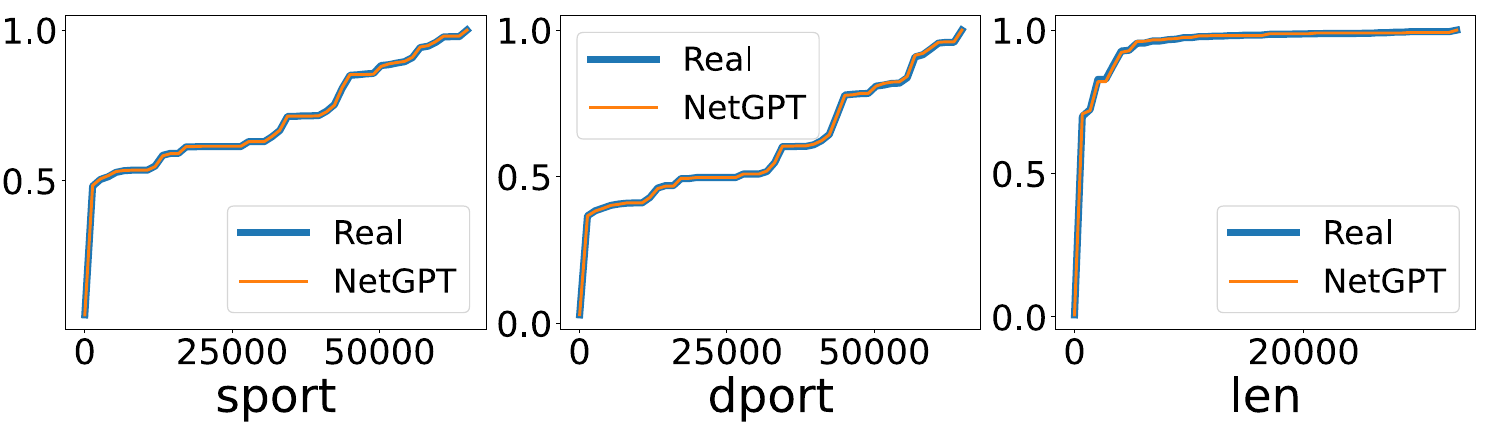}
	\end{minipage}}
	\subfigure[CDF on DoHBrw]{
	\begin{minipage}[t]{0.49\linewidth}
	\centering
	\includegraphics[width=1\textwidth]{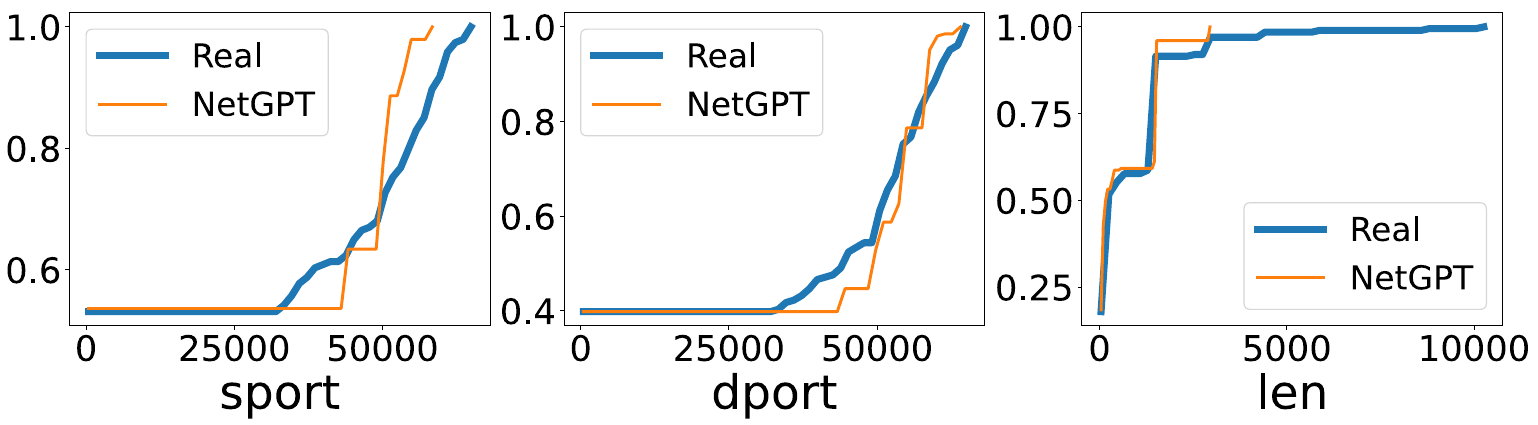}
	\end{minipage}}
	
	\subfigure[CDF on USTCTFC]{
	\begin{minipage}[t]{0.49\linewidth}
	\centering
	\includegraphics[width=\textwidth]{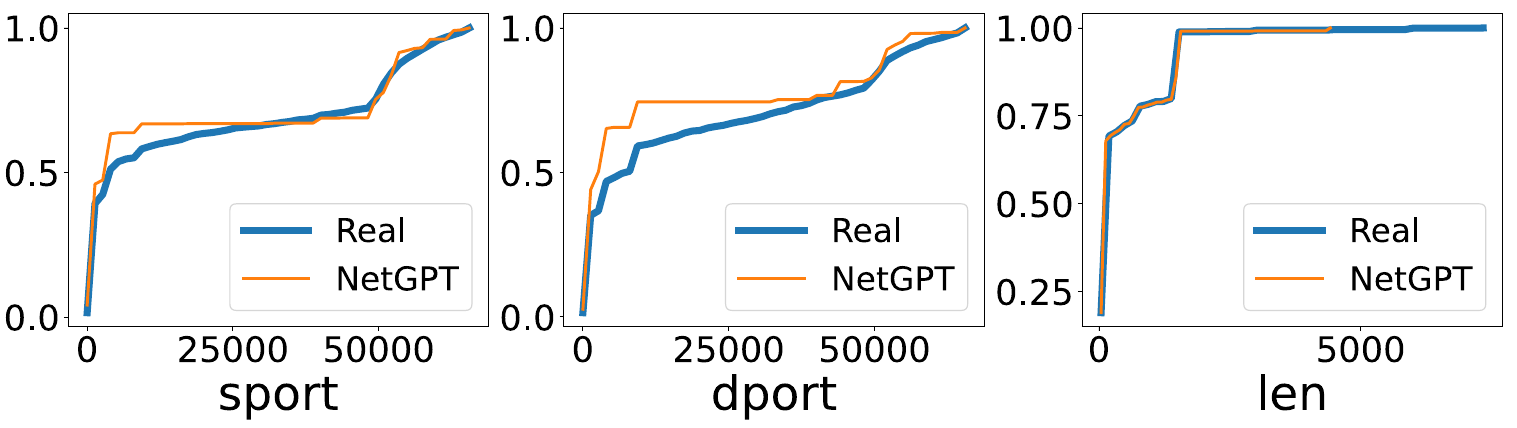}
	\end{minipage}}
	\subfigure[CDF on Cybermining]{
	\begin{minipage}[t]{0.49\linewidth}
	\centering
	\includegraphics[width=\textwidth]{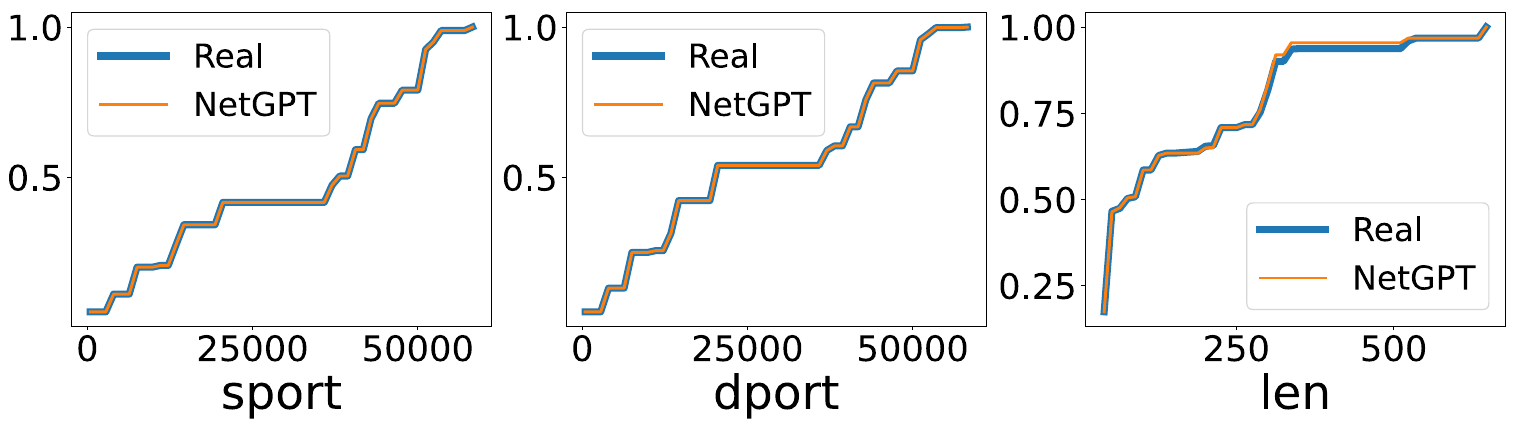}
	\end{minipage}}
	\centering
	\vskip -1em
	\caption{Distribution of NetGPT's generated header fields on four datasets.}
	\label{fig:cdf}
\end{figure*}

\begin{figure}[]
\centering
\includegraphics[width=0.45\textwidth]{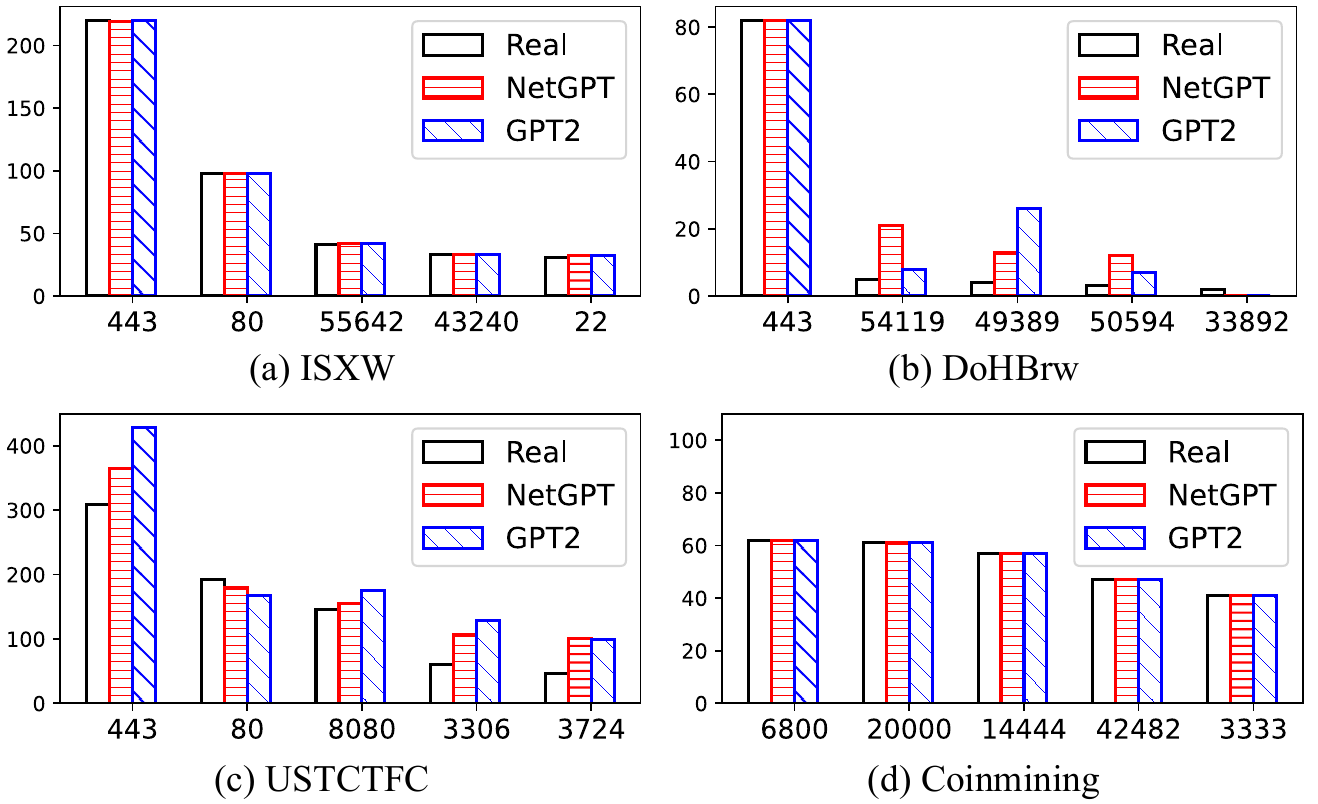}
\vskip -0.5em
\caption{Top 5 destination ports.}
\vskip -0.5em
\label{fig:dport}
\end{figure}

\subsection{Evaluation on Traffic Understanding} 
In this section, we conduct experiments to evaluate the performance of traffic understanding tasks. We present both packet-level and flow-level results, and all baselines and our NetGPT are fine-tuned for each task. Additionally, we provide NetGPT-A which is fine-tuned for all tasks simultaneously with different prompts.

From Table \ref{tb:TU}, we can make the following observations. \textbf{(1)} Generally, all these works can achieve excellent performance even in tasks 7 and 8 with an unseen dataset Cybermining, demonstrating the effectiveness of pretraining without considering downstream tasks. \textbf{(2)} Our NetGPT gains about 4 points on average from flow-level perspectives. Although all baselines perform much better on packet-level tasks than flow-level ones, our NetGPT can decrease much lower, indicating the effectiveness of segmenting packets in flows. \textbf{(3)} ET-BERT outperforms GPT-2 in most tasks from both packet-level and flow-level perspectives due to bidirectional learning that involves more semantic information to improve understanding performance. Nevertheless, in task 6, GPT-2 receives a much higher F1 than ET-BERT, which is because its text2text way is not affected by fixed class numbers. \textbf{(4)} Despite utilizing one-directional learning, 
NetGPT achieves competitive performance compared to ET-BERT and even gains much better performance in packet-level perspective for  task 6 and task 8 due to the effectiveness of shuffling header fields and text2text inference. \textbf{(5)} NetGPT-A and NetGPT show similar good performance, demonstrating that prompts can further help support multi-task learning.

\begin{table*}[]
\centering
\caption{Ablation study on traffic understanding performance.}
\label{tb:abU}
\scalebox{0.75}{
\begin{tabular}{c|cc|cc|cc|cc|cc|cc|cc|cc|c}
\toprule
\multirow{2}{*}{model} & \multicolumn{2}{c|}{task 1}          & \multicolumn{2}{c|}{task 2}          & \multicolumn{2}{c|}{task 3}          & \multicolumn{2}{c|}{task 4}         & \multicolumn{2}{c|}{task 5}          & \multicolumn{2}{c|}{task 6}          & \multicolumn{2}{c|}{task 7}          & \multicolumn{2}{c|}{task 8}  & Avg        \\ \cline{2-18} 
                       & \multicolumn{1}{c|}{AC}     & F1     & \multicolumn{1}{c|}{AC}     & F1     & \multicolumn{1}{c|}{AC}     & F1     & \multicolumn{1}{c|}{AC}    & F1     & \multicolumn{1}{c|}{AC}     & F1     & \multicolumn{1}{c|}{AC}     & F1     & \multicolumn{1}{c|}{AC}     & F1     & \multicolumn{1}{c|}{AC}     & F1  &-   \\ \hline \hline
                       
\textbf{NetGPT}              & \multicolumn{1}{c|}{0.9375} & 0.9352 & \multicolumn{1}{c|}{0.9683} & 0.8056 & \multicolumn{1}{c|}{1.0000} & 1.0000 & \multicolumn{1}{c|}{0.8120} & 0.7747 & \multicolumn{1}{c|}{1.0000} & 1.0000 & \multicolumn{1}{c|}{0.9563} & 0.9463 & \multicolumn{1}{c|}{1.0000} & 1.0000 & \multicolumn{1}{c|}{1.0000} & 1.0000 & 0.9460\\ \hline
w.o. Shuffle           & \multicolumn{1}{c|}{0.9375} & 0.9352 & \multicolumn{1}{c|}{0.9603} & 0.8021 & \multicolumn{1}{c|}{1.0000} & 1.0000 & \multicolumn{1}{c|}{0.8440} & 0.8267 & \multicolumn{1}{c|}{1.0000} & 1.0000 & \multicolumn{1}{c|}{0.9563} & 0.9460  & \multicolumn{1}{c|}{1.0000} & 1.0000 & \multicolumn{1}{c|}{1.0000} & 1.0000 & 0.9505\\ \hline
w.o. Segment               & \multicolumn{1}{c|}{0.9375} & 0.9352 & \multicolumn{1}{c|}{0.9286} & 0.7351 & \multicolumn{1}{c|}{1.0000} & 1.0000 & \multicolumn{1}{c|}{0.8280} & 0.8091 & \multicolumn{1}{c|}{1.0000} & 1.0000 & \multicolumn{1}{c|}{0.9504} & 0.9363 & \multicolumn{1}{c|}{1.0000} & 1.0000 & \multicolumn{1}{c|}{1.0000} & 1.0000 & 0.9413\\ \bottomrule
\end{tabular}}
\end{table*}

\begin{table}
\centering
\caption{Further ablation study.}
\scalebox{0.75}{
\begin{tabular}{c|cc|cc|cc}
\toprule
\multirow{2}{*}{model} & \multicolumn{2}{c|}{10 epochs} & \multicolumn{2}{c|}{50 epochs} & \multicolumn{2}{c}{100 epochs} \\ \cline{2-7} 
                       & \multicolumn{1}{c|}{AC}       & F1       & \multicolumn{1}{c|}{AC}        & F1       & \multicolumn{1}{c|}{AC}        & F1        \\ \hline\hline
\textbf{NetGPT}              & \multicolumn{1}{c|}{0.8320}   & 0.8028   & \multicolumn{1}{c|}{0.8120}    & 0.7747   & \multicolumn{1}{c|}{0.8080}    & 0.7676    \\ \hline
w.o. Shuffle           & \multicolumn{1}{c|}{0.7440}   & 0.6591   & \multicolumn{1}{c|}{0.8440}    & 0.8267   & \multicolumn{1}{c|}{0.8200}    & 0.7894    \\ \hline
w.o. Segment               & \multicolumn{1}{c|}{0.8280}   & 0.7973   & \multicolumn{1}{c|}{0.8280}    & 0.8091   & \multicolumn{1}{c|}{0.8480}    & 0.8325    \\ \bottomrule
\end{tabular}}
\label{tb:further}
\end{table}

\subsection{Evaluation on Traffic Generation} 
In this section, we conduct experiments to evaluate the performance of traffic generation tasks. We generate important header fields as described in Section \ref{sec:data}. To evaluate the fidelity of generated traffic, the JSD comparing with real traffic is presented in Table \ref{tb:TG}, and we also provide the AC and F1 results of generation in Appendix Table \ref{tb:TG1}. As more diverse IPs may provide better privacy, we do not show the JSD on IPs in Table \ref{tb:TG}, and evaluate their Diversity Ratio in Table \ref{tb:TG1}. Since packets in the same flow can have different header fields, we only conduct packet-level traffic generation. Also, as BERT-based models lack the generation ability, we do not compare with ET-BERT.

As lower JSD means better fidelity, from Table \ref{tb:TG}, we can make the following observations.
\textbf{(1)} Both NetGPT and GPT-2 can obtain great fidelity with much lower JSD around 0.1, especially compared to the existing GAN-based model with JSD \cite{Yin0JFS22} around [0.1,0.6]. \textbf{(2)} NetGPT
achieves the best average performance and outperforms GPT-2 in most datasets, demonstrating the effectiveness of shuffling in the traffic generation tasks. Specially, although GPT-2 has already gained great fidelity with low JSD, the JSD of NetGPT further decreases by about 50\% on ISXW.

We also confirm visually that the structure of these distributions greatly matches the original raw trace in Figure  \ref{fig:cdf}. Additionally, accurately capturing the distribution of header fields with large support show higher importance for downstream tasks, e.g., attack detection \cite{Yin0JFS22}. Figure \ref{fig:dport} shows NetGPT can better capture the structure of top-K destination ports. Results of source ports and length are also presented in Appendix Figure \ref{fig:sport} and \ref{fig:len}.

\subsection{Ablation Study} 
To evaluate the impact of  each module on downstream tasks, we conduct an ablation study in this section. For traffic generation tasks, as they are packet-level tasks for diverse header fields, we can only remove the shuffling module, which is actually the GPT-2 we trained. Since we have already provided a detailed analysis of NetGPT-A, NetGPT and GPT-2 in Table \ref{tb:TG}, we will not repeat it here.

For traffic understanding tasks, as we have already presented a detailed analysis of NetGPT-A and NetGPT in Table \ref{tb:TU}, we will also not repeat it here. In addition, as shown in Table \ref{tb:abU}, we compare NetGPT with two variants, namely, NetGPT without Shuffle, and NetGPT without Segment. Note that to include the comparison with NetGPT without Segment, we perform this experiment from a flow-level perspective. 
Our results show that NetGPT outperforms the other two variants in most tasks, indicating that both shuffling and segmenting have a positive impact. However, in task 4, NetGPT receives lower results, leading to lower average performance. We conduct further analysis as shown in Table \ref{tb:further}. Interestingly, with fewer epochs of training, NetGPT receives better performance even when compared to the results reported in Table \ref{tb:abU}. This may be because shuffling and segmenting may add complexity during finetuning and lead to overfitting with more epochs. Additionally, compared to NetGPT without Segment, NetGPT without Shuffle performs better in all tasks, which suggests that segmenting packets within a flow is more important than bidirectional information.